\documentstyle[amssymb,twocolumn,prl,aps]{revtex}
\begin{document}

%\draft

\twocolumn[
\hsize\textwidth\columnwidth\hsize\csname@twocolumnfalse\endcsname

\title{Parametric Excitation of Alfv\'{e}n Waves by Gravitational
Radiation}
\author{M. Servin, G.Brodin and M. Bradley}
\address{Department of Plasma Physics,
Ume{\aa} University,
SE-901 87 Ume{\aa}, Sweden}
\author{M. Marklund}
\address{Department of Electromagnetics,
Chalmers University of Technology,
SE-412 96 G\"oteborg, Sweden}
\maketitle

\begin{abstract}
We consider the parametric excitation of Alfv\'{e}n waves by gravitational radiation propagating on a Minkowski background, parallel to an external magnetic field. As a starting point, standard ideal MHD equations incorporating the curvature of space-time has been
derived. The growth rate of the Alfv\'{e}n waves has been calculated, using the normal mode approach. Various astrophysical applications of our investigations are discussed, and finally we demonstrate that the coupling coefficients of the interacting modes fulfill the Manley-Rowe relations.
\end{abstract}


]

\section{Introduction}

Interaction of electromagnetic fields with gravitational radiation has been
studied by several authors [1-19]. Besides a purely theoretical interest in
such phenomena, there is a number of different applications. For example in
astrophysics [1-5], in cosmology [6] and under laboratory conditions [7-10],
where - in the later case - the goal is to find suitable mechanisms to
detect gravitational radiation. Furthermore, there are many examples of
gravitational wave interaction that may take place in plasmas. This has been
studied during the eighties by a group at the Kazan School of gravitation
(see for example [1,10,11] and references therein) and more recently by
Refs. 4, 12, 13 and 14.

In Ref. 12 it was shown that parametric excitation of {\it high frequency}
plasma waves by gravitational radiation may take place. Due to the frequency
matching conditions, however, the plasma must be very thin for that process
to be possible, and the amount of energy transfer is therefore limited. In
the present paper we will thus consider parametric excitation of low
frequency MHD waves by gravitational waves, which - in contrast - may take
place in a comparatively dense plasma. The relevance of this problem for the
conversion of gravitational wave energy to the plasma inside supernovas has
previously been discussed by Ref. 14. However, due to the complexity of the
physical situation, a highly idealized model will be studied, where a one
dimensional monochromatic gravitational wave - superimposed on a flat
background metric - propagates through a homogeneous two component plasma.

The organization of the paper is as follows: In section II\ idealized
MHD-equations incorporating the effects of the gravitational wave are
derived, starting from covariant two-fluid equations. In section III
parametric excitation of shear Alfv\'{e}n and magnetosonic waves are
considered, the three wave coupling coefficients are derived and the growth
rate is found. By adding a phenomenological resistivity to the equations,
the threshold value of the gravitational amplitude is also calculated.
Finally, in section IV, our results are summarized and theoretical
considerations like energy conservation properties and the fulfillment of
Manley-Rowe relations as well as possible applications are discussed.

\section{Relativistic MHD-Equations}

\label{MHDeq}In order to obtain general relativistic fluid equations
governing a plasma we begin by considering a system consisting of a charged
perfect fluid and an electromagnetic field [20]. Introducing the restframe
scalar quantities: mass density (or rather energy density times $1/c^{2}$) $%
\rho _{(m)}$, charge density $\rho _{(q)}$, pressure $p$ and the 4-velocity
field (or fluid velocity) $u^{\mu }\equiv dx^{\mu }/d\tau $ and 4-current
density $j^{\mu }\equiv \rho _{(q)}u^{\mu }$, where $\tau $ is the proper
time and $x^{\mu }$ coordinates in the lab frame, this system is
characterized by having the energy-momentum tensor $T^{\mu \nu
}=T_{(fl)}^{\mu \nu }+T_{(em)}^{\mu \nu }$, where 
\begin{eqnarray*}
T_{(fl)}^{\mu \nu } &\equiv &\left( \rho _{(m)}+\frac{p}{c^{2}}\right)
u^{\mu }u^{\nu }-pg^{\mu \nu } \\
T_{(em)}^{\mu \nu } &\equiv &\frac{1}{\mu _{0}}\left( F^{\mu \tau }F_{\tau
}^{\ \nu }-\frac{1}{4}g^{\mu \nu }F^{\tau \sigma }F_{\sigma \tau }\right) 
\end{eqnarray*}
and $F_{\mu \nu }$ is the electromagnetic field tensor satisfying Maxwell's
equations 
\begin{equation}
F_{\ \ ;\nu }^{\mu \nu }=-\mu _{0}j^{\mu }  \label{eq:ME1}
\end{equation}
\begin{equation}
F_{\mu \nu ;\sigma }+F_{\nu \sigma ;\mu }+F_{\sigma \mu ;\nu }=0
\label{eq:ME2}
\end{equation}
We have adopted the convention that Greek suffixes $\mu ,\nu ,..$ have the
range $0,1,2,3$ and $i,j,..$ have the range $1,2,3$ and the metric tensor $%
g^{\mu \nu }$ has the signature (+ - - - ). \newline
\indent
The conservation laws of the system follows from that the 4-divergence of
the energy-momentum tensor vanishes, i.e. $T_{\ \ ;\mu }^{\mu \nu }=0$, and
with the use of Maxwell's equations one gets 
\begin{equation}
(\rho _{(m)}u^{\mu })_{;\mu }+\frac{p}{c^{2}}u_{\ ;\mu }^{\mu }=0
\label{eq:masscons}
\end{equation}
\begin{equation}
(\rho _{(m)}+\frac{p}{c^{2}})u^{\mu }u_{\ ;\mu }^{\nu }=(g^{\mu \nu }-\frac{1%
}{c^{2}}u^{\mu }u^{\nu })p_{,\mu }+F_{\ \mu }^{\nu }j^{\mu }
\label{eq:enrgmomcons}
\end{equation}
where eq.(\ref{eq:masscons}) is obtained by projection along the 4-velocity $%
u^{\mu }$. This equation is identified as energy balance in the rest-frame
of matter and gives the equation of continuity (mass conservation) in the
non relativistic limit. Equation (\ref{eq:enrgmomcons}) gives for $\nu =0$
the energy balance (modulo the content of (\ref{eq:masscons}), i.e., energy
balance in the non relativistic limit) and for $\nu =1,2,3$ momentum
balance. \newline
\indent Under the conditions of low internal energy the fluid description of
a plasma can be simplified by putting, for each particle species of the
plasma, $\rho _{(m)}=mn$ where $n$ is the restframe particle density and $m$
is the particle mass. Also without these restrictions we may put $\rho
_{(q)}=qn$ where $q$ is the particle charge. Suppose we have a plasma
consisting of two species of particles oppositely charged (i.e. $%
q_{1}/q_{2}=-1$), but, in general, with distinct masses. For each species we
assign a fluid satisfying the equations (\ref{eq:masscons}) and (\ref
{eq:enrgmomcons}). The fluids are assumed interpenetrating and interacting
through the electromagnetic field and, in general, the gravitational field.
We neglect the effect of particle collisions. If we assume non-relativistic
pressure, i.e. such that $mn\gg p/c^{2}$, and non-relativistic fluid
velocities - in the sense that we may neglect quadratic terms in $1/c$ in
the $\nu \neq 0$ components of eq.(\ref{eq:enrgmomcons}) - then we have, for
each of the two fluids, equations for conservation of particles (or mass)
and momentum in the form 
\begin{eqnarray*}
(nu^{\mu })_{;\mu } &=&0 \\
mnu^{\mu }u_{;\mu }^{i} &=&g^{ij}p_{,j}+F_{\ \mu }^{i}qnu^{\mu }
\end{eqnarray*}
Maxwell's equations remain the same if we let $j_{\mu }$ be the total
current density.

Under the conditions that for both species $\partial _{t}\sim \omega \ll
\omega _{c}$ and $C_{A}^{2}\ll c^{2},$ this two-fluid description can be
cast into a set of single-fluid equations. By $\partial _{t}\sim \omega $ we
mean that a characteristic frequency, $\omega $, can be assigned to the time
variations in the dynamical quantities [21]. We use $\omega _{c}\equiv
|q|B/m $ for the cyclotron frequency, $B$ is the magnetic field intensity
(which can be obtained from the Lorentz frame components of the
electromagnetic field tensor) and $C_{A}$ is the Alfv\'{e}n velocity,
defined by $C_{A}^{2}=B^{2}/\mu _{0}mn$. Under the above conditions it
follows that the two fluid velocities are approximately equal and that the
particle densities may be regarded as exactly equal. Furthermore, we let the
equation of state be the one of isothermal compression and assume $|u^{\mu
}u_{\ ;\mu }^{i}|\ll |F_{\ 0}^{i}qnu^{0}|/mn\propto |F_{\ j}^{i}qnu^{j}|/mn$
(meaning that the electric and magnetic forces approximately balance each
other). In light of this, we obtain the following set of single fluid
equations (MHD-equations) 
\begin{equation}
(nu^{\mu })_{;\mu }=0  \label{eq:SFec}
\end{equation}
\begin{equation}
(m_{(1)}+m_{(2)})nu^{\mu }u_{\ ;\mu }^{i}=g^{ij}p_{,j}+F_{\ \mu }^{i}j^{\mu }
\label{eq:SFem}
\end{equation}
\begin{equation}
F_{\ \mu }^{i}u^{\mu }=-\frac{1}{nq_{(1)}}g^{ij}\left( \frac{%
m_{(2)}p_{(1),j}-m_{(1)}p_{(2),j}}{m_{(1)}+m_{(2)}}\right)  \label{eq:SFol}
\end{equation}
\begin{equation}
{p}_{,i}=k_{B}{T}n_{,i}  \label{eq:SFes}
\end{equation}
Equation (\ref{eq:SFec}) and (\ref{eq:SFem}) are obtained by adding the two
particle conservation equations and the two equations of momentum balance,
respectively, and setting the velocities equal (when added), letting $j$ be
the total current density, $p$ the total pressure and $T\equiv
T_{(1)}+T_{(2)}.$ The suffixes ${}_{(1)}$ and ${}_{(2)}$ refers to the two
distinct particle species. By subtracting the two equations of momentum
conservation, where the terms $u^{\mu }u_{\ ;\mu }^{i}$ have been neglected,
one finds eq. (\ref{eq:SFol}), which we refer to as the generalized Ohm's
law.

Note that if the right hand side of eq. (\ref{eq:SFol}) is negligible as
compared to $F_{\ 0}^{\imath }u^{0}$ and $F_{\ j}^{\imath }u^{j}$ this
equation simply reads $F_{\ \mu }^{i}u^{\mu }=0$ and we can then refer to
the single-fluid equations as the ideal MHD-equations. In the following we
limit ourselves to this case. Note that the MHD-equations are not
independent of Maxwell's and Einstein's field equations, since we used
Maxwell's equations and $T_{\ \ ;\mu }^{\mu \nu }=0$, that follows from
Einstein's equations, in deriving them.

We now consider gravitational radiation on a Minkowski background treating
the plasma as a testfluid. Thus, the plasma back scattering effect on the
gravitational field is lost. The gravitational radiation is chosen to be
weak gravitational waves in the transverse traceless (TT) gauge propagating
in the $x^{3}$-direction. This plane wave solution of the linearized
Einstein field equations can be written 
\begin{eqnarray*}
ds^{2} &=&c^{2}dt^{2}-(1-h_{+})dx^{2}-(1+h_{+})dy^{2} \\
&&+2h_{\times }dxdy-dz^{2}
\end{eqnarray*}
where $h\equiv \tilde{h}e^{ik_{\mu }x^{\mu }}+{\rm c.c.}$ and $|\tilde{h}%
|\ll 1$ with the wave vector $[k^{\mu }]=(\omega /c,0,0,k)$ satisfying the
dispersion relation $k^{\mu }k_{\mu }=0$. In all the following calculations
we neglect terms that are quadratic in $\tilde{h}$ or higher. The metric
tensor can thus be written as $g_{\mu \nu }=\eta _{\mu \nu }+h_{\mu \nu }$
where $\eta _{\mu \nu }$ is the metric tensor of Minkowski space and $h_{\mu
\nu }$ represents the small, $|h_{\mu \nu }|\ll 1$, fluctuation in the
gravitational field.

The nonzero Christoffel symbols are then calculated to 
\begin{eqnarray*}
\Gamma _{\ 01}^{1} &=&-\Gamma _{\ 02}^{2}=\Gamma _{\ 11}^{0}=-\Gamma _{\
22}^{0}=-\Gamma _{\ 13}^{1} \\
&=&\Gamma _{\ 23}^{2}=\Gamma _{\ 11}^{3}=-\Gamma _{\ 22}^{3}=\frac{1}{2}\dot{%
h}_{+}
\end{eqnarray*}
\[
\Gamma _{\ 02}^{1}=\Gamma _{\ 01}^{2}=-\Gamma _{\ 23}^{1}=-\Gamma _{\
13}^{2}=\Gamma _{\ 12}^{3}=\Gamma _{\ 12}^{0}=\frac{1}{2}\dot{h}_{\times } 
\]
where $\dot{h}\equiv \partial h/\partial \xi $ and $\xi \equiv x^{3}-x^{0}$.

By expanding the covariant derivative, eq.(\ref{eq:SFec} ) becomes $(nu^{\mu
})_{,\mu }=0$ as a result of $\Gamma _{\ \nu \mu }^{\mu }u^{\nu }=0$ and -
noting that $g^{\mu \nu }=\eta ^{\mu \nu }-h^{\mu \nu }$ to first order,
such that $g^{ij}=-\delta ^{ij}-h^{ij}$ - eq.(\ref{eq:SFem}) reads 
\begin{equation}
mnu^{\mu }u_{\ ,\mu }^{i}+G^{i}=-\delta ^{ij}p_{,j}-h^{ij}p_{,j}+F_{\ \mu
}^{i}j^{\mu }
\end{equation}
where we have introduced $G^{i}=mn\Gamma _{\ \nu \mu }^{i}u^{\nu }u^{\mu }$
and $m=m_{(1)}+m_{(2)}$ .

Next, we perform the same expansion in Maxwell's equations and rewrite them
in terms of the electromagnetic field tensor in the form $F_{\ \nu }^{\mu }$%
. The idea is to express all field tensor terms in the same form, preferably
the one that gives the most simple expressions. We can separate $F_{\ \ ;\nu
}^{\mu \nu }=-\mu _{0}j^{\mu }$ into two equations. Setting $\mu =0$ we
obtain a Poisson-like equation which we discard - since in the MHD-regime $%
j^{0}\approx 0$. Setting $\mu =i$ we read off ''Ampere's law``, 
\begin{equation}
\delta ^{jk}F_{\ k,j}^{i}=\mu _{0}j^{i}-(h^{j\nu }F_{\ \nu
}^{i})_{,j}+\Gamma _{\ \tau j}^{i}g^{j\nu }F_{\ \nu }^{\tau }+\Gamma _{\
\tau j}^{j}g^{\tau \nu }F_{\ \nu }^{i}
\end{equation}
From eq.(\ref{eq:ME2}), which by symmetry in $F^{\mu \nu }$ is equivalent to 
$F_{\mu \nu ,\sigma }+F_{\nu \sigma ,\mu }+F_{\sigma \mu ,\nu }=0$, we
obtain a number of trivial identities, a generalized equation for $\nabla
\cdot {\bf B}$ and ''Faraday's law``: 
\begin{eqnarray*}
F_{\ 0,2}^{3}-F_{\ 2,3}^{0}+F_{\ 3,0}^{2} &=&(h_{\times }F_{\text{ }%
3}^{1}-h_{+}F_{\text{ }3}^{2})_{,0} \\
-F_{\ 0,1}^{3}+F_{\ 1,3}^{0}-F_{\ 3,0}^{1} &=&-(h_{+}F_{\text{ }%
3}^{1}+h_{\times }F_{\text{ }3}^{2})_{,0} \\
F_{\ 0,1}^{2}-F_{\ 1,2}^{0}+F_{\ 2,0}^{1} &=&(h_{+}F_{\text{ }%
2}^{1}+h_{\times }F_{\text{ }2}^{2})_{,0} \\
&&+(h_{\times }F_{\text{ }0}^{1}-h_{+}F_{\text{ }0}^{2})_{,1}
\end{eqnarray*}
For notational purposes it is convenient to introduce an abstract basis $\{%
{\bf \hat{x}},{\bf \hat{y},\hat{z}}\}$. The one-fluid equations and the
Maxwell's equations above can then be written in a vector representation
with an algebraic structure identical to the Euclidean. We define 
\begin{eqnarray*}
{\bf x} &\equiv &x{\bf \hat{x}}+y{\bf \hat{y}}+z{\bf \hat{z}}\equiv x^{1}%
{\bf \hat{x}}+x^{2}{\bf \hat{y}}+x^{3}{\bf \hat{z}} \\
{\bf v} &\equiv &v_{x}{\bf \hat{x}}+v_{y}{\bf \hat{y}}+v_{z}{\bf \hat{z}}%
\equiv u^{1}{\bf \hat{x}}+u^{2}{\bf \hat{y}}+u^{3}{\bf \hat{z}} \\
{\bf j} &\equiv &j_{x}{\bf \hat{x}}+j_{y}{\bf \hat{y}}+j_{z}{\bf \hat{z}}%
\equiv j^{1}{\bf \hat{x}}+j^{2}{\bf \hat{y}}+j^{3}{\bf \hat{z}} \\
{\bf E} &\equiv &E_{x}{\bf \hat{x}}+E_{y}{\bf \hat{y}}+E_{z}{\bf \hat{z}}%
\equiv cF_{\ 1}^{0}{\bf \hat{x}}+cF_{\ 2}^{0}{\bf \hat{y}}+cF_{\ 3}^{0}{\bf 
\hat{z}} \\
{\bf B} &\equiv &B_{x}{\bf \hat{x}}+B_{y}{\bf \hat{y}}+B_{z}{\bf \hat{z}}%
\equiv F_{\ 3}^{2}{\bf \hat{x}}+F_{\ 1}^{3}{\bf \hat{y}}+F_{\ 2}^{1}{\bf 
\hat{z}} \\
\nabla  &\equiv &\partial _{x}{\bf \hat{x}}+\partial _{y}{\bf \hat{y}}%
+\partial _{z}{\bf \hat{z}}\equiv \frac{\partial \ }{\partial x^{1}}{\bf 
\hat{x}}+\frac{\partial \ }{\partial x^{2}}{\bf \hat{y}}+\frac{\partial \ }{%
\partial x^{3}}{\bf \hat{z}}
\end{eqnarray*}
(Note that these quantities differ to first order in $\tilde{h}$ from what
an observer in the lab system would measure.) One then obtains the following
set of equations governing the plasma 
\begin{eqnarray}
mn\left( \partial _{t}{\bf v}+({\bf v}\cdot \nabla ){\bf v}\right)  &=&{\bf j%
}\times {\bf B}-\nabla p+{\bf g}  \label{eq:MHDmom} \\
{\bf E}+{\bf v}\times {\bf B} &=&0  \label{eq:MHDohm} \\
\partial _{t}n+\nabla \cdot (n{\bf v}) &=&0  \label{eq:MHDcont} \\
\nabla p &=&k_{B}T\nabla n  \label{eq:MHDstate} \\
{\bf \nabla \times B} &=&\mu _{0}{\bf j+j}_{E}  \label{eq:rotB} \\
{\bf \partial }_{t}{\bf B+\nabla \times E} &=&{\bf -j}_{B}  \label{eq:rotE}
\end{eqnarray}
where 
\begin{eqnarray*}
{\bf g} &=&-{\bf G}+{\bf g}_{{\rm pressure}}+{\bf g}_{{\rm em-coupling}} \\
{\bf G} &=&-mn\left[ (v_{z}-c)(\dot{h}_{+}v_{x}+\dot{h}_{\times
}v_{y}),\right.  \\
&&\left. (v_{z}-c)(\dot{h}_{\times }v_{x}-\dot{h}_{+}v_{y}),\right.  \\
&&\left. \frac{1}{2}\dot{h}_{+}(v_{x}^{2}-v_{y}^{2})-\dot{h}_{\times
}v_{x}v_{y}\right]  \\
{\bf g}_{{\rm pressure}} &=&-\left[ h_{+}\partial _{x}p+h_{\times }\partial
_{y}p,h_{\times }\partial _{x}p-h_{+}\partial _{y}p,0\right]  \\
{\bf g}_{{\rm em-coupling}} &=&\left[ {\bf -}h_{\times
}B_{z}j_{x}-(h_{+}B_{y}-h_{\times }B_{x})j_{z},\right.  \\
&&\left. 2h_{+}B_{z}j_{x}+h_{\times }B_{z}j_{y},-(h_{+}B_{x}+h_{\times
}B_{y})j_{y}\right]  \\
{\bf j}_{E} &\equiv &\left[ \partial _{y}(h_{+}B_{z})+\partial
_{z}(h_{+}B_{y}-h_{\times }B_{x}),\right.  \\
&&\left. -\partial _{x}(h_{+}B_{z}),-\partial _{x}(h_{+}B_{y}-h_{\times
}B_{y})\right]  \\
{\bf j}_{B} &\equiv &\left[ \partial _{t}(h_{+}B_{x}+h_{\times
}B_{y}),0,-\partial _{t}(h_{+}B_{z})\right] 
\end{eqnarray*}
{\bf \ }We want to point out that terms of the order $E\dot{h}_{\times }/c$
in{\bf \ }${\bf j}_{E}${\bf \ }have been neglected, since $E/B$ is of the
order $C_{A}$ in ideal MHD theory. {\bf \ }Furthermore, note that $cF_{\
0}^{1}\neq cF_{\ 1}^{0}\equiv E_{x}$, etc., which is the origin of some of
the terms appearing in the expressions for ${\bf g}_{{\rm em-coupling}}$, $%
{\bf j}_{E}$ and ${\bf j}_{B}$. In addition to eqs.(\ref{eq:MHDmom})-(\ref
{eq:rotE}) Maxwell's equations produce constraints (e.g. for $\nabla \cdot 
{\bf B}$), however it is easy to verify that these constraints are
propagated by the equations of time evolution, (\ref{eq:MHDmom}), (\ref
{eq:MHDcont}) and (\ref{eq:rotE}).

\section{Wave-wave interactions}

From now on we will consider resonant three-wave interaction between
gravitational radiation and MHD-waves. Such processes may occur whenever
quadratic nonlinear terms, such as the right hand sides of eqs. (\ref
{eq:MHDmom})-(\ref{eq:rotE}), are present, and the dispersion relations of
the interacting waves allow for the frequency and wave-vector matching
conditions to be fulfilled. For a general review of resonant three wave
interaction in plasmas, see e.g. Ref. 22.

In the absence of any waves we assume to have the configuration of a static
homogeneous, $n=n^{(0)}$, magnetized, ${\bf B}={\bf B}^{(0)},$ plasma in
Minkowski space. Cartesian coordinates are chosen ($x^{\mu }=(ct,x,y,z)$)
for a frame in which the velocity field (and the current density field)
vanishes. The gravitational waves are then inferred as small perturbations
to the Minkowski background, as in the previous section, and the MHD-waves
as the existence of the small fluctuations: $n^{(1)},{\bf v}^{(1)},{\bf j}%
^{(1)},{\bf E}^{(1)},{\bf B}^{(1)}$. Furthermore, in order to simplify the
algebra, we make the assumptions that the direction of ${\bf B}^{(0)}$ is
everywhere parallel to the direction of propagation of the gravitational
waves, i.e. ${\bf B}^{(0)}=B^{(0)}{\bf \hat{z}}$, and that the gravitational
radiation is polarized such that $h_{+}=0$.

\subsection{Linear Calculations}

It is instructive to first investigate the linearized theory in some detail.
Linearizing the equations (\ref{eq:MHDmom})-(\ref{eq:rotE}) in the variables 
$h_{\times },n^{(1)},{\bf v}^{(1)},{\bf j}^{(1)},{\bf E}^{(1)},{\bf B}^{(1)}$
we find that the gravitational waves do not drive plasma perturbations
linearly. This is a consequence of the direction of propagation of the
gravitational wave (parallel to the magnetic field) that was chosen.
Similarly the linear plasma perturbations are the ordinary MHD-modes.
Fourier analyzing we obtain the dispersion relations for the shear Alfv\'{e}%
n wave 
\begin{equation}
D_{A}\equiv \omega ^{2}-C_{A}^{2}k_{z}^{2}=0  \label{eq:DRa}
\end{equation}
and for the fast and slow magnetosonic wave 
\begin{equation}
D_{m}\equiv \omega ^{4}-\omega
^{2}k^{2}(C_{S}^{2}+C_{A}^{2})+k_{z}^{2}k^{2}C_{S}^{2}C_{A}^{2}=0.
\label{eq:Drms}
\end{equation}
The constants introduced are the Alfv\'{e}n velocity $C_{A}^{2}\equiv {%
B^{(0)}}^{2}/{mn^{(0)}\mu _{0},}$ the thermal velocity $C_{S}^{2}\equiv
k_{B}T/m$, and we have used the notation ${\bf k}=k_{x}{\bf \hat{x}}+k_{z}%
{\bf \hat{z}}$ together with $k=\left| {\bf k}\right| $. In the next
subsection we will consider superposition of MHD-waves, and by expressing
all variables in terms of the fluid velocity we can represent the solution
as a sum of eigenvectors 
\begin{equation}
\left( 
\begin{array}{cc}
n^{(1)} &  \\ 
{\bf v}^{(1)} &  \\ 
{\bf j}^{(1)} &  \\ 
{\bf E}^{(1)} &  \\ 
{\bf B}^{(1)} & 
\end{array}
\right) =\sum_{\alpha }\left( 
\begin{array}{ccc}
\theta _{\alpha }n^{(0)} &  &  \\ 
{\bf v}_{\alpha }^{(1)} &  &  \\ 
\frac{i\theta _{\alpha }}{\mu _{0}}{\bf k}_{\alpha }\times {\bf B}^{(0)}-%
\frac{i\sigma _{\alpha }}{\mu _{0}}{\bf k}_{\alpha }\times {\bf v}_{\alpha
}^{(1)} &  &  \\ 
{\bf B}^{(0)}\times {\bf v}_{\alpha }^{(1)} &  &  \\ 
\theta _{\alpha }{\bf B}^{(0)}-\sigma _{\alpha }{\bf v}_{\alpha }^{(1)} &  & 
\end{array}
\right)   \label{eq:eigenv}
\end{equation}
where $\theta _{\alpha }\equiv {\bf k}_{\alpha }\cdot {\bf v}_{\alpha
}^{(1)}/\omega _{\alpha }$, $\sigma _{\alpha }\equiv k_{\alpha
z}B^{(0)}/\omega _{\alpha }$ and $\alpha $ is a wave-mode index. \newline
\indent
As we intend to study nonlinear wave coupling it is convenient to adopt the
normal mode method of approach [22], which typically simplifies the algebra
in the nonlinear stage of the calculations. We define a normal mode as a
linear combination, $a_{\alpha }$, of the dynamical quantities that to
linear order satisfies 
\begin{equation}
\partial _{t}a_{\alpha }+i\omega _{\alpha }a_{\alpha }=0  \label{eq:nm}
\end{equation}
The dynamical quantities are now only assumed to have harmonic spatial
dependence, i.e. $\nabla =i{\bf k}_{\alpha }$. From eq. (\ref{eq:nm}) the
proper linear combinations are 
\begin{equation}
a_{A}=v_{y}^{(1)}-\frac{\omega _{A}}{{k_{A}}_{z}B^{(0)}}B_{y}^{(1)}
\label{eq:Amode1}
\end{equation}
for the Alfv\'{e}n mode, and 
\begin{eqnarray}
a_{m} &=&n^{(1)}+\chi v_{x}^{(1)}+\frac{n^{(0)}{k_{m}}_{z}}{\omega _{m}}%
v_{z}^{(1)}  \nonumber \\
&&-\frac{{k_{m}}_{z}C_{A}^{2}\chi }{B^{(0)}\omega _{m}}B_{x}^{(1)}+\frac{{%
k_{m}}_{x}C_{A}^{2}\chi }{B^{(0)}\omega _{m}}B_{z}^{(1)}  \label{eq:MSmode1}
\end{eqnarray}
for the magnetosonic modes, with the frequency-wave number pairs $(\omega
_{A},{\bf k}_{A})$ and $(\omega _{m},{\bf k}_{m})$ satisfying the dispersion
relations $D_{A}=0$ and $D_{m}=0$, respectively. The constant $\chi $ is
defined as $\chi \equiv n_{0}(\omega _{m}^{2}-C_{S}^{2}{k_{m}}%
_{z}^{2})/C_{S}^{2}{k_{m}}_{x}\omega _{m}$. With aid of the corresponding
eigenvectors and the relation $v_{z}^{(1)}=v_{x}^{(1)}C_{S}^{2}{k_{m}}_{x}{%
k_{m}}_{z}/(\omega _{m}^{2}-C_{S}^{2}{k_{m}^{2}}_{x})$ we can, after some
algebraic manipulations, write the normal modes as 
\begin{eqnarray}
a_{A} &=&2v_{y}^{(1)}=\frac{1}{\omega _{A}}\frac{\partial D_{A}}{\partial
\omega _{A}}v_{y}^{(1)}  \label{eq:Amode2} \\
a_{m} &=&\frac{n^{(0)}}{\omega _{m}^{2}C_{S}^{2}{k_{m}^{2}}_{x}}\frac{%
\partial D_{m}}{\partial \omega _{m}}{v_{x}^{(1)}}.  \label{eq:MSmode2}
\end{eqnarray}
For the nonlinear calculation we need the eigenvectors expressed in terms of
the normal modes, and the final linear results are 
\begin{equation}
\left( 
\begin{array}{cc}
n^{(1)} &  \\ 
{\bf v}^{(1)} &  \\ 
{\bf j}^{(1)} &  \\ 
{\bf E}^{(1)} &  \\ 
{\bf B}^{(1)} & 
\end{array}
\right) _{A}=a_{A}\left( 
\begin{array}{ccc}
0 &  &  \\ 
(0,\frac{1}{2},0) &  &  \\ 
(\frac{iB^{(0)}{k_{A}^{2}}_{z}}{2\omega _{A}\mu _{0}},0,-\frac{iB^{(0)}{k_{A}%
}_{x}{k_{A}}_{z}}{2\omega _{A}\mu _{0}}) &  &  \\ 
(-\frac{B^{(0)}}{2},0,0) &  &  \\ 
(0,-\frac{B^{(0)}{k_{A}}_{z}}{2\omega _{A}},0) &  & 
\end{array}
\right)   \label{eq:eA}
\end{equation}
\newline
and 
\begin{equation}
\left( 
\begin{array}{cc}
n^{(1)} &  \\ 
{\bf v}^{(1)} &  \\ 
{\bf j}^{(1)} &  \\ 
{\bf E}^{(1)} &  \\ 
{\bf B}^{(1)} & 
\end{array}
\right) _{m}=a_{m}c_{m}\left( 
\begin{array}{ccc}
1+\frac{C_{S}^{2}{k_{m}^{2}}_{z}}{\omega _{m}^{2}-C_{S}^{2}{k_{m}^{2}}_{z}}
&  &  \\ 
(\zeta ,0,\zeta \frac{C_{S}^{2}{k_{m}}_{x}{k_{m}}_{z}}{\omega
_{m}^{2}-C_{S}^{2}{k_{m}^{2}}_{z}}) &  &  \\ 
(0,-\frac{iB^{(0)}k_{m}^{2}}{n^{(0)}{k_{m}}_{x}\mu _{0}},0) &  &  \\ 
(0,\frac{\omega _{m}B^{(0)}}{n^{(0)}{k_{m}}_{x}},0) &  &  \\ 
(-\frac{{k_{m}}_{z}B^{(0)}}{n^{(0)}{k_{m}}_{x}},0,\frac{B^{(0)}}{n^{(0)}}) & 
& 
\end{array}
\right)   \label{eq:em}
\end{equation}
where $c_{m}\equiv \omega _{m}^{2}C_{S}^{2}{k_{m}^{2}}_{x}/2(\omega
_{m}^{4}-k_{m}^{2}{k_{m}^{2}}_{z}C_{A}^{2}C_{S}^{2})$ and $\zeta \equiv
\omega _{m}/n^{(0)}{k_{m}}_{x}$

\subsection{Nonlinear Calculations}

The aim of this section is to investigate the lowest order nonlinear
influence of the gravitational radiation on the MHD modes described above.
In particular we are interested in the threshold value (for parametric
excitation) of the gravitational amplitude, and the growth rates of the
excited MHD waves. We will again assume that the wave vectors lies in $xz$%
-plane, i.e. ${\bf k}=k_{x}{\bf \hat{x}}+k_{z}{\bf \hat{z}}$ for the
MHD-waves, but in contrast to the case of linear wave modes this is a
restriction made in order to simplify the algebra [23].

We consider coherent three-wave interactions, the three waves being one
gravitational wave and two MHD-waves, with the matching conditions 
\begin{eqnarray}
\omega _{g} &=&\omega _{I}+\omega _{II}  \label{Freqmatch} \\
{\bf k}_{g} &=&{\bf k}_{I}+{\bf k}_{II}  \label{vectormatch}
\end{eqnarray}
where $I$ and $II$ are indexing the MHD-waves. In the nonlinear regime the
normal modes does no longer satisfy eq.(\ref{eq:nm}), but rather 
\[
\partial _{t}a_{\alpha }+i\omega _{\alpha }a_{\alpha }=\left( [\partial
_{t}a_{\alpha }]_{{\rm n.l}.}\right) _{{\bf k}_{\alpha }}
\]
where ${\rm n.l}.$ denotes (first order) nonlinear terms and the suffix ${%
{\bf k}_{\alpha }}$ indicates that terms not oscillating as $e^{i{\bf k}%
\cdot {\bf x}}$ vanishes due to rapid oscillations. Explicit forms for the
right hand side is found by using the original expressions for the normal
modes - eqs. (\ref{eq:Amode1}) or (\ref{eq:MSmode1}) - together with eqs. (%
\ref{eq:MHDmom})-(\ref{eq:rotE}).

We let index $I$ denote the magnetosonic wave perturbation, index $II$ the
Alfv\'{e}n wave perturbation and we use a complex representation (i.e.
letting $f\rightarrow f+f^{\ast }$ for all variables, where the star denotes
complex conjugate). Making use of the linear eigenvectors (\ref{eq:eA}) and (%
\ref{eq:em}) as approximations in the nonlinear right hand sides, we obtain
the coupled mode equations 
\begin{eqnarray}
\partial _{t}a_{I}+i\omega _{I}a_{I} &=&C_{I}a_{II}^{\ast }h_{\times }
\label{CME1} \\
\partial _{t}a_{II}+i\omega _{II}a_{II} &=&C_{II}a_{I}^{\ast }h_{\times }
\label{CME2}
\end{eqnarray}
after lengthy but straightforward algebra, where the coupling coefficients
are [24] 
\begin{eqnarray}
C_{I} &=&-\frac{i}{2}\frac{n_{0}k_{xI}}{\omega _{I}^{2}-C_{A}^{2}k_{I}^{2}}%
\omega _{I}\omega _{g}  \label{eq:C1} \\
C_{II} &=&-\frac{i}{2}\frac{1}{n_{0}}\frac{\omega _{I}^{2}C_{S}^{2}k_{xI}}{%
(\omega _{I}^{4}-C_{S}^{2}C_{A}^{2}k_{I}^{2}k_{zI}^{2})}\omega _{II}\omega
_{g}  \label{eq:C2}
\end{eqnarray}
In deriving the expressions for $C_{I}$ and $C_{II}$ we have also applied $%
{\bf k}_{I}\approx -{\bf k}_{II}$, which follows from the matching condition
(\ref{vectormatch}) together with $C_{A}\ll c$ and the dispersion relations.

From eq. (\ref{eq:C1}) and (\ref{eq:C2}) one may get the incorrect
impression that the coupling strength diverges in the limit $%
C_{S}^{2}\rightarrow 0$. Thus in order to shed some light on our formulas in
the cold limit, we first renormalize 
\[
\begin{array}{lll}
a_{I} & \rightarrow  & C_{S}^{2}a_{I} \\ 
C_{I} & \rightarrow  & C_{S}^{2}C_{I} \\ 
C_{II} & \rightarrow  & C_{II}/C_{S}^{2}
\end{array}
\]
and then take the limit $C_{S}^{2}\rightarrow 0$. The corresponding coupling
coefficients then becomes 
\begin{eqnarray}
C_{I} &=&-\frac{i}{2}\frac{n_{0}\omega _{I}}{k_{xI}}\omega _{g} \\
C_{II} &=&-\frac{i}{2}\frac{k_{xI}\omega _{II}}{n_{0}\omega _{I}^{2}}\omega
_{g}
\end{eqnarray}
Another special case of particular interest is the limit of parallel
propagation (but with arbitrary ratio $C_{S}^{2}/C_{A}^{2}$), in which case
the magnetosonic dispersion relation coincides with that of the shear Alfv%
\'{e}n wave, and the only distinction between the modes is the polarization,
which differ by 90 degrees. Again the general coupling coefficient (\ref
{eq:C1}) seem to diverge, since from the magnetosonic dispersion relation $%
\omega _{I}^{2}-C_{A}^{2}k_{I}^{2}\sim k_{xI}^{2}$ when $k_{xI}^{2}%
\rightarrow 0$. However, by using another renormalization 
\[
\begin{array}{lll}
a_{I} & \rightarrow  & a_{I}/\chi  \\ 
C_{I} & \rightarrow  & C_{I}/\chi  \\ 
C_{II} & \rightarrow  & \chi C_{II}
\end{array}
\]
and taking the limit ${k_{I}}_{x},{k_{II}}_{x}\rightarrow 0$, we obtain 
\begin{equation}
C_{I}=C_{II}=-\frac{i}{2}\omega _{g}  \label{CP}
\end{equation}
Since dissipation of the waves have not been included in our model, the
instability threshold value of the gravitational amplitude found from (\ref
{CME1}) and (\ref{CME2}) is so far zero. However, since only weak
dissipation is of interest we can take such effects into account by simply
substituting $\partial _{t}a_{\alpha }\rightarrow (\partial _{t}+\gamma
_{\alpha })a_{\alpha }$ in the coupled mode equations [25], where $\gamma
_{\alpha }$ is the linear damping rate of the mode $\alpha $. The most
common damping mechanism of MHD waves is that due to finite resistivity.
Calculating the linear damping by replacing (\ref{eq:MHDohm}) with ${\bf E}+%
{\bf v}\times {\bf B}=\eta {\bf j}$, where $\eta $ is the resistivity, we
find $\gamma _{\alpha }=\eta k_{\alpha }^{2}/\mu _{0}$. Next we introduce
the (weakly time dependent) normal mode amplitudes, $A_{\alpha }$, defined
by $a_{\alpha }=A_{\alpha }e^{-i\omega _{\alpha }t}$, where $\alpha =I,II$.
Substituting these expressions into (\ref{CME1}) and (\ref{CME2}) taking the
damping into account, we find the general form for the condition of
parametric growth of waves $|\widetilde{h}_{\times }|>h_{{\rm thr}}\equiv
(\gamma _{I}\gamma _{II}/C_{I}C_{II}^{\ast })^{1/2}$[22], where $\widetilde{h%
}_{\times }$ is the amplitude of the gravitational wave and $h_{{\rm thr}}$
is the threshold value for parametric excitation. In the limit of parallel
propagation we find from (\ref{CP}) that the threshold value $h_{{\rm thr}}$
reduces to 
\begin{equation}
|\widetilde{h}_{\times }|>h_{{\rm thr}}\approx \frac{4\gamma _{I},_{II}}{%
\omega _{g}}=\frac{\eta \omega _{g}}{\mu _{0}C_{A}^{2}}
\end{equation}
Furthermore, if the gravitational amplitude is well above threshold ($|%
\widetilde{h}_{\times }|\gg h_{{\rm thr}}$) the general expression for the
parametric growth rate $\Gamma $ from (\ref{CME1}) and (\ref{CME2}) is $%
\Gamma \approx \sqrt{C_{I}C_{II}^{\ast }}|\widetilde{h}_{\times }|$ [22],
and the result for the special case of parallel propagation is$\newline
$%
\begin{equation}
\Gamma \approx \frac{1}{2}\omega _{g}|\widetilde{h}_{\times }|\newline
\label{eq:growth}
\end{equation}
It should be pointed out that in addition to the wave interactions
considered above, we have found zero coupling coefficients for a number of
cases. To be specific: For the same polarization of the gravitational pump
wave ($h_{+}=0$), and propagation parallel to the external magnetic field,
the following combinations of MHD-waves {\em cannot} be excited in the
resonant three wave approximation, since the coupling coefficient then
becomes zero: 1) Two ion-acoustic (or slow magnetosonic) modes. 2) One
ion-acoustic and one Alfv\'{e}n wave. 3) Two Alfv\'{e}n waves with the same
linear polarization.

Note that for the case of non-zero coupling considered above, the Alfv\'{e}n
waves have perpendicular polarizations in the parallel limit. The coupling
between the differently polarized modes then results from the quadratic
nonlinear terms proportional to $\dot{h}_{\times }v_{y}{\bf \hat{x}}$ and $%
\dot{h}_{\times }v_{x}{\bf \hat{y}}$ in eq. (\ref{eq:MHDmom}), and from
similar cross terms in (\ref{eq:rotE}) that couple the two different
polarizations through $h_{\times }$. The dependence of the results on the
various polarizations for parallel propagation can be physically understood
as follows: For the MHD-waves to gain energy from the gravitational wave,
the MHD-waves must be able to reduce the gravitational wave amplitude.
Including the source term in the wave equation for the gravitational wave,
we immediately see that it is only the component $T_{xy}$ of the energy
momentum tensor that may affect the gravitational wave with $h_{\times }$%
-polarization. However, for the three ''null-cases'' listed above, it is
trivial to see that $T_{xy}$ vanishes (within the quadratic MHD
approximation), and thus the gravitational wave is unaffected by the
presence of such waves. From energy conservation [26]{\bf \ }it is thus
clear that the corresponding coupling must disappear. For perpendicular
polarization of the Alfv\'{e}n waves, on the other hand, $%
T_{xy}=mn_{0}v_{x}v_{y}-B_{x}B_{y}/\mu _{0}\neq 0$, and thus the
gravitational amplitude is affected by such a combination of waves and in
accordance with this we have found the coupling to be non-zero.

\section{Summary and Discussion}

We have considered parametric excitation of Alfv\'{e}n waves by
gravitational radiation propagating parallel to the external magnetic field.
As a starting point, standard ideal MHD equations (i.e. without special
relativistic effects) incorporating the curvature of space time have been
derived. It should be pointed out that the system of equations (\ref{eq:SFec}%
) -(\ref{eq:SFes}) in principle can be used in situations where we have
strong deviation from Minkowski space time, although the condition of
non-relativistic fluid velocities then limits the applicability. Focusing on
the case where the metric is that of a small amplitude monochromatic
gravitational wave superimposed on flat space-time, the growth rate for
nonlinearly coupled shear Alfv\'{e}n and fast magnetosonic waves have been
found.

In our calculations we have considered a monochromatic gravitational pump
wave, which could be produced by binary systems. As seen from (\ref
{eq:growth}) (or more generally from (\ref{eq:C1}) and (\ref{eq:C2}) which
applies for arbitrary directions of propagations [27]), the growth rate is
roughly of the order $\Gamma \thicksim h_{\times }\omega _{g}$. Thus the
plasma parameters $n_{0}$, $B_{0}$ and $T$ do not significantly influence
the growth rate, at least not as long as the assumptions of the derivation
is fulfilled. It is not hard to find a plasma fulfilling these assumptions,
e.g. choosing $n_{0}\thicksim 10^{14}$ ${\rm m}^{-3}$[28] and considering
waves with frequency $\omega \lesssim 10^{4}$ ${\rm rad}$ ${\rm s}^{-1}$, $%
B_{0}$ may attain any value roughly in the interval $10^{-5}-1$ ${\rm T}$
for the estimate $\Gamma \thicksim h_{\times }\omega _{g}$ to apply, where
the limits of the interval comes from the conditions $\omega \ll $ $\omega
_{c}$ and $C_{A}^{2}\ll c^{2}$ respectively. Furthermore, the condition $%
C_{S}^{2}\ll c^{2}$ still allows for comparatively high plasma temperatures,
since it is obviously fulfilled provided the thermal velocities of the
particles are much smaller than the speed of light. As for the source of
gravitational radiation we follow Ref. 12 and consider a binary system of
two equal masses $m=3M_{\odot }$ separated by a distance of six
Schwarzschild radii $r=12Gm/c^{2}$ emitting gravitational radiation with
frequency of the order $\omega _{g}\approx 10^{4}$ rad/s. By considering two
point masses separated by a fixed distance $r$ the amplitude $\tilde{h}%
_{\times }$ at a distance $R$ from the system is estimated to give $\tilde{h}%
_{\times }\sim Gmr^{2}\omega _{g}^{2}/(2c^{4}R)$ . Also in the case of
Ref.12, which considered parametric excitation of high-frequency plasma
waves, the growth rate fulfilled $\Gamma \sim h_{\times }\omega _{g}$,
implying the growth rate $\Gamma \sim 10^{-2}\ {\rm s}^{-1}$ at a distance
of 1/60 {\rm au} from the source, where a process at a closer distance was
ruled out by the frequency matching conditions combined with the linear
dispersion relations. In our case the linear dispersion relations and
matching conditions allow a parametric process closer to the source, and
thereby opens up the possibility for a higher growth rate, although too
close to the source the background plasma may be too inhomogeneous and too
far from steady state for our calculations to be applicable.

Furthermore, excitation of MHD waves may take place in a dense plasma, and
therefore processes such as supernovas are of interest, where gravitational
wave absorption may take place inside the exploding star. In a discussion of
possible mechanisms of absorbing gravitational wave energy in supernovas
Ref. 14 has written ''Since the effect of acceleration by gravitational
waves is independent of mass of the charge, both the ions and the electron
respond in an identical manner, which is not the case for electromagnetic
waves. This means that waves such as Alfv\'{e}n waves which describe
oscillation of charge neutral plasmas are ideal. The coupling, however, is
weak.'' At the present stage of understanding it is too early to deduce
whether significant gravitational wave absorption by MHD waves may occur.
Calculations taking into account the effects of a broad band gravitational
spectrum, plasma inhomogeneities, etc., must first be performed. In
particular inhomogeneity scale lengths with a scale length significantly
shorter than the wavelength of the gravitational mode - such as at the
plasma boundary of the supernova - may lead to excitation of MHD surface
waves with a significantly enhanced growth rate as compared to the present
homogeneous plasma coupling mechanism. This is in analogy with parametric
excitation scenarios for high frequency plasma surface waves [29], where the
surface waves may have a considerably higher growth rate than the
corresponding bulk waves, provided the inhomogeneity scale length is
considerably shorter than the wave length of the pump wave. Such a problem,
however, is a project for future work.

An interesting question from a theoretical point of view is whether the
coupling coefficients $C_{I}$ and $C_{II}$ of eqs. (\ref{eq:C1}) and (\ref
{eq:C2}) satisfy the Manley-Rowe relations [22]. Generally such relations
follow from an underlaying Hamiltonian structure of the governing equations,
and assures that each of the decay products takes energy from the pump wave
in direct proportion to their respective frequencies. This means that the
parametric process can be interpreted quantum mechanically - i.e. we can
think of a three wave process as the decay of a pump wave quanta with energy 
$\hbar \omega _{g}$ into wave quantas with energy $\hbar \omega _{I}$ and $%
\hbar \omega _{II}$ respectively. An interesting consequence is that
generally a lot of three wave decay processes are forbidden from the start
by the Manley--Rowe relations (for example the decay of a plasmon into two
photons), since they imply that we only get a positive growth rate when the
pump wave has the highest frequency, in consistence with the quantum picture
[30]. It should be pointed out that all well established basic systems of
equations in plasma physics such as the ideal MHD-equations, the
Vlasov-Maxwell equations and the standard multi-fluid equations, all possess
the underlaying Hamiltonian structure that leads to fulfillment of the
Manley-Rowe relations for three-wave interaction processes [31]. However,
since the theory of gravitation differs in important respects from other
fundamental theories of physics, it is an open question whether the
Manley-Rowe relations applies also when one of the interacting modes is a
gravitational wave. To investigate this issue we consider the energy
increase rate of mode I. Multiplying Eq. (\ref{CME1}) with $v_{xI}^{\ast
}mk_{I}^{2}C_{s}^{2}(\omega _{I}-k_{zI}^{2}C_{A}^{2})/\omega _{I}k_{xI}$ and
using the eigenvectors (\ref{eq:eA}) and (\ref{eq:em}) we find 
\begin{equation}
\frac{\partial W_{I}}{\partial t}=\omega _{I}V  \label{W1}
\end{equation}
where $V=2%
%TCIMACRO{\func{Im}}%
%BeginExpansion
\mathop{\rm Im}%
%EndExpansion
[mn_{0}\omega _{g}h_{\times }v_{xI}^{\ast }v_{yII}^{\ast }/\omega _{I}]$ and
the magnetosonic wave energy density is $W_{I}=mn_{0}k_{I}^{2}(\omega
_{I}-k_{zI}^{2}C_{A}^{2})(\partial D_{m}/\partial \omega _{I})\left|
v_{xI}\right| ^{2}/2k_{xI}^{2}\omega _{I}$ (cf. Eq. (20) in Ref 32).
Similarly, multiplying Eq. (\ref{CME2}) with $mn_{0}v_{yII}^{\ast }$, and
applying (\ref{eq:eA}) and (\ref{eq:em}) we obtain

\begin{equation}
\frac{\partial W_{II}}{\partial t}=\omega _{II}V  \label{W2}
\end{equation}
where $W_{II}=mn_{0}\left| v_{yII}\right| ^{2}$ (cf. Eq. (13) in Ref 32). Up
to now, we have treated the gravitational wave as a pump wave, i.e. we have
only considered the initial stage of an instability where the energy density
of the daughter waves are small enough for the influence on the
gravitational wave to be neglected. For a practical purpose this regime may
apply until nonlinear saturation mechanisms (outside our calculation scheme)
sets in for the growing MHD-waves, unless the plasma is extremely dense.
From a theoretical point of view, however, it is of interest to consider the
decrease in the gravitational wave amplitude due to the growth of the MHD
perturbations. Including the energy momentum source term in the wave
equation for the gravitational wave, i.e.using $\square h_{\mu \nu }=$ $%
-2\kappa T_{\mu \nu }\equiv -16\pi GT_{\mu \nu }/c^{4}$, keeping only the
resonantly varying part of $T_{\mu \nu }$ (i.e. the part proportional to $%
\exp [{\rm i}(k_{g}z-\omega _{g}t)]$) and letting $h_{12}=\widetilde{h}%
_{\times }(t)\exp [{\rm i}(k_{g}z-\omega _{g}t)]+{\rm c.c.}$, we immediately
obtain 
\begin{equation}
{\rm i}\omega _{g}\frac{\partial \widetilde{h}_{\times }}{\partial t}=\kappa
(T_{12})_{{\bf k}_{g}}=\kappa \left[ mn_{0}v_{xI}v_{yII}-\frac{B_{xI}B_{yII}%
}{\mu _{0}}\right] 
\end{equation}
which, after using the eigenvectors (\ref{eq:eA}) and (\ref{eq:em}) and
multiplying with $\omega _{g}\widetilde{h}_{\times }^{\ast }$ can be written 
\begin{equation}
\frac{\partial W_{g}}{\partial t}=-\omega _{g}V  \label{W3}
\end{equation}
where $W_{g}=\omega _{g}^{2}|\widetilde{h}_{\times }|^{2}/2\kappa $ is the
energy density of the gravitational wave [33]. Since the same factor $V$
appears in Eqs (\ref{W1}), (\ref{W2}) and (\ref{W3}), the Manley--Rowe
relations are indeed fulfilled (i.e. each mode changes energy in direct
proportion to its frequency), and furthermore we see that the total wave
energy $W=W_{I}+W_{II}+W_{g}$ is conserved, which follows from (\ref{W1}), (%
\ref{W2}) and (\ref{W3}) together with the frequency matching condition (\ref
{Freqmatch}).


\begin{references}
\bibitem{Restnote}  Naturally the condition $\partial /\partial t\ll \omega
_{c}$ cannot hold in arbitrary reference frames. We assume it to be true in
the rest frame of the fluid, but due the condition of non-relativistic fluid
velocities it also holds in all frames of relevance.
\bibitem{Tetrad-note}  In order to check the algebra we have verified that
identical coupling coefficents $C_{I}$ and $C_{II}$ follows from a tetrad
formalism.
\bibitem{Note 4}  Generally it is not entirely unproblematic to make energy
conservation arguments in general relativity. However, as will be clear from
the last section, energy conservation using the standard expression for the
gravitational wave energy can be applied to our case.
\bibitem{Note 5}  Using $\Gamma =\sqrt{C_{I}C_{II}^{\ast }}h_{\times }\omega
_{g}$ and estimating $C_{I}$ and $C_{II}$ as given by (31) and (32),
applying the dispersion relations and the frequency matching conditions, we
get $\Gamma \sim h_{\times }\omega _{g}$ also for non-parallel propagation
and independently of the ratio $C_{S}^{2}/C_{A}^{2}.$
\bibitem{Note 6}  The possible particle densities in the example of
considerations have a very wide range. For example $n_{0}\sim 10^{4}{\rm m}%
^{-3}$ for low-density unperturbed interstellar matter. The value $n_{0}\sim
10^{14}{\rm m}^{-3}$chosen by us somewhat arbitrarily, is of the some order
of magnitude as the density in the solar corona. Much larger number
densities, as found in accretion discs, are also possible.
\end{references}
\end{document}